\documentclass[journal]{vgtc}

\usepackage{mathtools}  
\usepackage{footmisc}

\title{Multi-user Reset Controller for Redirected Walking\\Using Reinforcement Learning}

\author{
Ho Jung Lee, Sang-Bin Jeon, Yong-Hun Cho and In-Kwon Lee
}

\authorfooter{
  \item Ho Jung Lee, Sang-Bin Jeon, and In-Kwon Lee are with Department of Computer Science, Yonsei University. E-mail: {dearshawn | ludens0508 | iklee}@yonsei.ac.kr.
   \item Yong-Hun Cho is with Digital eXPerience Lab, Korea University, Seoul 03722, South Korea. Email: maxburst88@gmail.com
}
\abstract{The reset technique of Redirected Walking (RDW) forcibly reorients the user's direction overtly to avoid collisions with boundaries, obstacles, or other users in the physical space. However, excessive resetting can decrease the user’s sense of immersion and presence. Several RDW studies have been conducted to address this issue. Among them, much research has been done on reset techniques that reduce the number of resets by devising reset direction rules (e.g.,~ 2:1-turn, reset-to-center) or optimizing them for a given environment. However, existing optimization studies on reset techniques have mainly focused on a single-user environment. In a multi-user environment, the dynamic movement of other users and static obstacles in the physical space increase the possibility of resetting. In this study, we propose a multi-user reset controller (MRC) that resets the user taking into account both physical obstacles and multi-user movement to minimize the number of resets. MRC is trained using multi-agent reinforcement learning to determine the optimal reset direction in different environments. This approach enables MRC to effectively account for different environmental contexts, including arbitrary physical obstacles and the dynamic movements of other users in the same physical space. We compared MRC with other reset techniques through simulation tests and user studies, and our results show that MRC reduces the mean number of resets by up to 55\%. Overall, our study confirmed that MRC is an effective reset technique in multi-user environments. Supplemental materials are available at an anonymous link: (\url{https://osf.io/rpftu/?view_only=8230f344502f4013af2a5229db5e21c3}).
  
}

\keywords{Virtual reality, Redirected walking, Resetting, Reinforcement learning}

\teaser{
    \centering
    \subfloat[\label{1a}]{
    \includegraphics[width=0.3\linewidth]{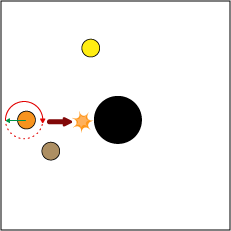}
    }
    \hfill
    \subfloat[\label{1b}]{
    \includegraphics[width=0.3\linewidth]{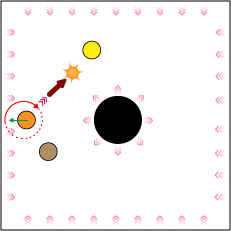}
    }
    \hfill
    \subfloat[\label{1c}]{
    \includegraphics[width=0.3\linewidth]{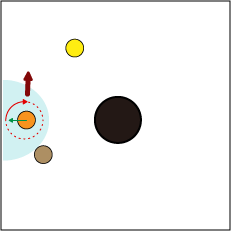}
    }
  \caption{Reset control during multi-user RDW experience in a physical space with a circle obstacle in the center: (a) Reset-to-Center (R2C): the user unconditionally reset to the center of the physical space without considering other users or obstacle in the center. (b) Reset-to-Gradient (R2G): the user reset to the direction of the sum of the force that is acted on them by the boundary of the physical space, obstacles, and other users. (c) Multi-user Reset Controller (MRC): the user reset to the optimal direction determined by a model trained with information about the user's physical space.}
  \label{fig:resetcontroller}
}
\renewcommand{\manuscriptnotetxt}{}
\graphicspath{{figs/}{Paper/figures/}{pictures/}{images/}{./}} 

\usepackage{tabu}                      
\usepackage{booktabs}                  
\usepackage{lipsum}                    
\usepackage{mwe}                       
\usepackage{mathptmx}                

\begin{document}

\firstsection{Introduction}

\maketitle

Virtual Reality (VR) is growing in popularity with advances in Head Mounted Display (HMD) devices. These devices allow people to enjoy VR content that is difficult to experience in the real world, such as gaming, training, and communication, which would provide a high sense of presence when walking through the interior of a virtual world while wearing an HMD \cite{usoh1999walking}. However, walking without manipulation is tricky due to the physical difference between physical and virtual space.

Redirected Walking (RDW) method proposed by Razzaque et al.~\cite{razzaque2005redirected} allows users to experience infinite walking in a virtual space from a limited physical space by manipulating the mapping ratio of the user's movement in the physical space to the virtual space imperceptibly (e.g.,~ subtle technique) or by manipulating the structure of the virtual environment (e.g.,~ change blindness). Walking with these RDW methods gives the user a higher sense of presence than using teleportation and joysticks \cite{steinicke2009estimation, steinicke2009real, langbehn2018evaluation}. However, although RDW methods use subtle techniques to adjust the user's trajectory in physical space, there are limitations in preventing collisions between the user and the boundaries or obstacles in physical space due to the physical space constraints. If the user is in danger of a collision, the user's HMD screen will display a command to stop moving and reorient the user to a walkable direction (e.g.,~ 2:1 Turn), which is called a reset \cite{williams2007exploring}. Frequent rests reduce the user's immersion and presence in the virtual reality experience \cite{suma2012taxonomy}. In addition, when users share a physical space with obstacles and experience VR, in addition to boundary reset, which reorients the user just before a collision with a wall or obstacle that is the boundary of the physical space, a user reset must also be performed to prevent collisions with other users in the VR experience. In the multi-user environment, the size of the walkable space is reduced, and more resets occur due to other dynamically moving users \cite{messinger2019effects, bachmann2019multi}. In this case, the loss of immersion can be minimized by using a reset technique that rotates users in the direction based on a specific rule or optimized for the environment \cite{suma2012taxonomy}.

By proposing different rules for the reset technique that determine the reset direction, studies have been conducted to reduce the number of resets \cite{williams2007exploring,thomas2019general,9881577,9733261}. In addition, a study has recently been proposed to optimize previous reset techniques with a local algorithm according to the shape of the physical space \cite{zhang2022adaptive}. However, we still need to explore a reset technique that determines the optimal reset direction for users in the presence of not only static obstacles but also dynamic obstacles, such as other users sharing and walking in the same physical space. In our research, we propose Multi-user Reset Controller (MRC) that determines the optimal reset direction to reduce the number of resets when two or more users experience RDW in the same physical space. In the case of Reset-to-Center (R2C) in \cref{1a}, the user will always reset to the center of the physical space, despite the presence of obstacles in the center of the physical space, and even if users use Reset-to-Gradient (R2G) in \cref{1b}, users cannot exclude the possibility of resetting in the direction that may collide with other users. In contrast, our MRC in \cref{1c} has an artificial intelligence model that determines the optimal reset direction at each moment.

MRC is trained through multi-agent reinforcement learning to observe dynamic user movements and reset users to the optimized directions as they experience irregular physical spaces with various obstacles. Our experiments show that MRC reduces the number of resets during the RDW experience by up to 55$\%$ compared to other reset techniques.

\section{Related Work}
\subsection{Redirected Walking}
RDW method proposed by Razzaque et al.~\cite{razzaque2005redirected} allows the user to explore the virtual space that is larger than the physical space. In recent years, research on RDW has been active and has become one of the key areas in virtual reality research \cite{nilsson201815, suma2012taxonomy}. RDW is divided into subtle and overt techniques \cite{suma2012taxonomy}. The subtle technique applies redirection gains within a detection threshold that is imperceptible to the user, creating a movement difference between physical and virtual space and avoiding collisions with boundaries. Steinicke et al.~\cite{steinicke2009estimation, steinicke2009real} measured the detection thresholds of the redirection gains because if the redirection gains are applied higher than the user's detection threshold, cyber-sickness with degraded immersion may occur. Since then, detection thresholds have been studied for various situations, and redirection gains within the range of these detection thresholds are applied in subsequent RDW studies \cite{cho2021walking, kim2021adjusting, williams2019estimation, li2021detection, langbehn2017bending}. The overt technique generates UI on the user's screen or plays sound effects to redirect the user. The reset technique is an example of an overt technique, which stops the user's movement and reorients the user when the user reaches the boundary or obstacle in the physical space with a possible risk of collision, even if the user's gait is adjusted by subtle techniques \cite{williams2007exploring}.

Redirection controllers, which combine subtle and overt techniques, have been actively researched recently \cite{nilsson201815}. Redirection controllers can be divided into reactive and predictive controllers. Reactive controllers consist of algorithms that redirect a user to a specific position or pattern, such as S2C \cite{razzaque2005redirected}, S2O \cite{razzaque2005redirected}, APF \cite{thomas2019general}, ARC \cite{williams2021arc}. On the other hand, predictive controllers predict the user's walking path and redirect the user in the optimal direction, such as FORCE \cite{zmuda2013optimizing}, MPCRed \cite{nescher2014planning}, S2OT \cite{lee2019real}, and SRL \cite{strauss2020steering}.

\subsection{Reset Technique}
Williams et al.~\cite{williams2007exploring} first propose a reset technique that stops the user's movement and reorients the user in a walkable direction. Freeze-Backup and Freeze-Turn freeze the user's HMD screen and command the user to move or turn in place to a walkable orientation, respectively. On the other hand, 2:1 Turn does not freeze the user's HMD screen, but uses rotation gain to reorient the user in place by rotating $180\,^{\circ}$ in the physical space and $360\,^{\circ}$ in the virtual space. The 2:1 Turn allows the user to continue walking in the direction of experience before resetting in the virtual space, whereas, in the physical space, the user would turn around and continue walking in the opposite direction to avoid colliding with a boundary. Based on the 2:1 Turn, which requires the lease amount of reset time and does not freeze the HMD screen, many reset techniques have been explored.

R2C \cite{azmandian2015physical, 5759437} modifies 2:1 Turn to reset the user to the center of physical space. The general assumption that the safest direction for the user to reset to is toward the center of the physical space works very well, despite its simplicity, if we don't consider the obstacles that are likely to be placed in the physical space. Later, Thomas et al.~\cite{thomas2019general} improved R2C to reset perpendicular to the boundary in situations where obstacles prevent resetting to the center of physical space. Thomas also proposed R2G, which resets in the direction of the sum of the forces acting on the user in an APF-enabled environment, and Step-Forward Reset to Gradient (SFR2G), which reorients the reset when the sum of the forces faces an obstacle or wall. Zhang et al.~\cite{zhang2022adaptive} proposed a local optimization algorithm to reduce the number of resets based on the previously proposed reset techniques (R2C, R2G, SFR2G) in several physical spaces. In addition to resets that rotate in place before colliding with a boundary, various methods have also been studied, such as out-of-place resetting (OPR) \cite{9733261}, which allows the users to redirect optimal physical locations to continue walking when performing a reset. Another method is to perform a reset even if there is no collision, to avoid performing a reset directly in front of a point of interest (POI) set as a target \cite{9881577}. However, to date, reset techniques have primarily been studied in a single-user environment.

\subsection{Multi-User Redirected Walking}
Azmandian et al.~\cite{azmandian2017evaluation} first proposed the concept of space sharing strategy and space partitioning strategy for multi-user RDW experience. Various multi-user redirection controllers have been proposed by applying the space sharing strategy. Bachman et al.~\cite{bachmann2019multi} extended APF to multi-user environment, and Messinger et al.~\cite{messinger2019effects} verified the performance when multi-users applied APF to RDW experience in physical spaces of different shapes and sizes. Lee et al.~\cite{lee2020optimal} proposed a predictive controller called MS2OT using a model trained by reinforcement learning. Jeon et al.~\cite{jeon2022dynamic} proposed OSP that learns to dynamically partition the physical space that is optimally served to multiple users through reinforcement learning. APF and MS2OT were designed as redirection controllers suitable for multi-user environments, but their reset controller is still based on earlier reset techniques (R2G, 2:1 Turn). OSP was trained to dynamically adjust the shutter so that the users reset in the proper direction. However, the range in which the shutters can move is limited by the range in which the topology of the shutters must be maintained in physical space, which limits the ability to reset the user in the optimal direction.

\begin{figure*}[t!]
    \centering
        \includegraphics[width=0.8\linewidth]{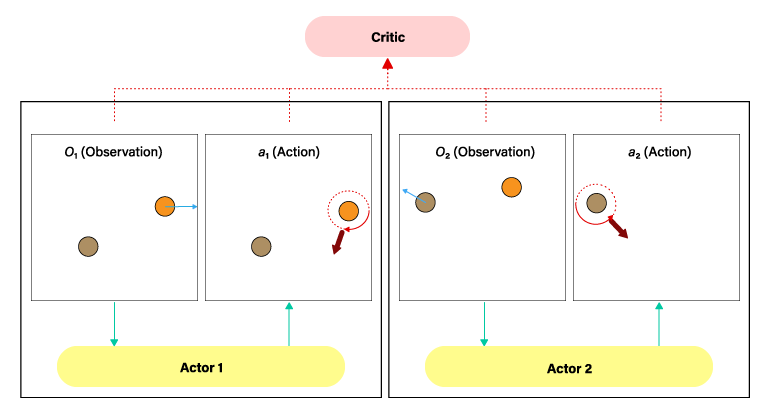} 
        \caption{Overview of MRC model: Following the CTDE method, critic collects the environment information observed by each user at each time a reset occurs, and determines the actions accordingly. MRC is trained to determine the action that maximizes the cumulative reward by calculating the reward for a given action. An actor corresponds to each user experiencing RDW. In the execution phase, actors corresponding to users according to the learned policy determine the optimal action based on the surrounding environment information.}
        \label{fig:model}
\end{figure*}

\subsection{Reinforcement Learning}
Reinforcement learning is a type of machine learning in which an agent learns to maximize its cumulative reward by performing the optimal action $a_t$ for a given state $s_t$ at a given time $t$ in a given environment \cite{sutton1998introduction}. Various reinforcement learning algorithms have been proposed in the reinforcement learning community, such as Proximal Policy Optimization (PPO) \cite{schulman2017proximal} and Soft Actor-Critic (SAC) \cite{haarnoja2018soft}. Redirection controllers trained with the proposed reinforcement learning algorithm are also being actively studied \cite{lee2019real, lee2020optimal, jeon2022dynamic, strauss2020steering, chang2021redirection}.

In an environment where multiple entities are interacting, each entity must cooperate for good performance. However, since single agent reinforcement learning determines actions without considering other entities in the environment, it may not be able to learn policy that cooperate with other entities. To overcome this problem, there is active research on Multi-Agent Reinforcement Learning (MARL), where multiple entities learn to cooperate with each other to achieve a common goal \cite{bucsoniu2010multi}. Among them, the Central Training and Decentralized Execution (CTDE) framework learns by collecting the information observed by all agents in the environment by a critic, and when the model is executed, it determines the action according to the local information observed by each actor \cite{lowe2017multi}. Because the CTDE paradigm learns from multiple agents sharing their policies, it can learn the optimal policy faster and more effectively than a single agent. Multi-Agent Posthmous Credit Assignment (MA-POCA) proposed by Cohen et al.~\cite{cohen2021use} performs well with a self-attention layer \cite{vaswani2017attention}, even when the number of agents changes during a learning episode. 

\section{Method}

\subsection{Reinforcement Learning based Reset Controller}
We assume that there are $n$ users in the current RDW. We design a Markov Decision Process (MDP) \cite{puterman2014markov} for the agents to learn reset directions that minimize the number of resets during the users' RDW experience. Since multiple users need to cooperate to achieve the common goal of reducing the number of resets during the experience, we use the MARL method to train MRC. Among them, by taking advantage of the MA-POCA structure of the CTDE method, the model learned from the two-user simulation as an episode is applied to the environment where two or more people experience RDW. We used the Unity ML-Agents Toolkit (ver. 2.2.1) \cite{juliani2018unity} to train MRC model by defining state, action, and reward functions. \cref{fig:model} shows an overview of MRC model. The simulation environment that is used for learning and experiment is implemented based on the Open-RDW library \cite{li2021openrdw}.

\subsection{States}
In MRC, the critic collects information about the physical space observed by each actor at each time step $t$, when one of the users is in a reset situation, and represents the state $s_t$ as follows:
    \begin{equation}
        s_t = (o_1, o_2, ... , o_{n}),
    \end{equation}
where $o_i$ is the physical space information observed by the actor corresponding to the $i$-th user at time step $t$. The information $o_{i}$ observed by the $i$-th actor is as follows:
    \begin{equation}
        o_{i} = (u_{i} , \theta_{i}),
        ~~i = 1, 2, ... , n,
    \end{equation}
where $u_{i}$ and $\theta_{i}$ are the physical 2D position and 1D orientation of the $i$-th user, respectively (\cref{fig:model}). Thus, the $i$-th actor corresponding to the $i$-th user observes $o_{i}$ with dimension $3$ at each time step $t$. The $u_{i}$ is scaled to $[-1, 1]$ relative to the size of the physical space, and the $\theta_{i}$ is scaled to $[-1, 1]$ relative to the arc degree. This allows MRC to be used in environments of different scales, even if they are trained on physical spaces of different sizes, as long as the shapes of the physical spaces are the same.
\begin{figure}[t!]
        \centering
        \begin{subfigure}[b]{0.4\columnwidth}
            \centering
            \includegraphics[width=\textwidth]{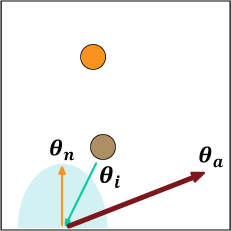}
            \caption{boundary reset case}
            \label{3a}
            \end{subfigure}
        \begin{subfigure}[b]{0.4\columnwidth}
            \centering
            \includegraphics[width=\textwidth]{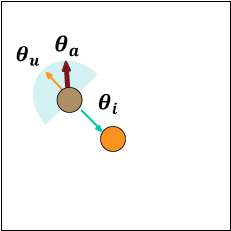}
            \caption{user reset case}
            \label{3b}
        \end{subfigure}
        \subfigsCaption{Reset process by action of MRC: $\theta_i$: user's current direction of movement, $\theta_n$: The normal direction of the wall or obstacle that caused the reset, $\theta_u$: The opposite direction of $\theta_i$, $\theta_a$: the reset direction determined by MRC.}
        \label{fig:action}
\end{figure}
\subsection{Action}
The goal of learning MRC is to determine the optimal reset direction based on the observed $s_t$ when a reset occurs in a multi-user RDW environment. \cref{fig:action} indicates a reset in the direction determined by MRC. Let $\theta_n$ be the normal direction of the wall or obstacle that caused the boundary reset (\cref{3a}). Let $\theta_u$ be the opposite of $\theta_i$, the direction of the $i$-th user who needs to do a user reset (\cref{3b}). To follow a consistent MDP for both types of resets, we define the action $a_t$ to the following reset direction $\theta_a$:
    \begin{equation}
    \begin{aligned}
        \theta_{a} \in \begin{cases}
        \left[\theta_{n} - \displaystyle \frac {\pi}{2}, \theta_{n} + \displaystyle \frac {\pi}{2}\right],~~ \mbox{if boundary reset},\\[10pt]
        \left[\theta_{u} - \displaystyle \frac {\pi}{2}, \theta_{u} + \displaystyle \frac {\pi}{2}\right],~~ \mbox{if user reset,}
        \end{cases}
    \end{aligned}
    \end{equation}
where $\theta_{a}$ is determined by the model learned so far, and is the optimal direction in which we can expect to avoid collisions with boundaries and other users as much as possible. The range of $\theta_a$ is calculated based on $\theta_n$ or $\theta_u$, depending on the type of reset. After performing a reset in this determined direction, the user can continue walking in the direction that best avoids collisions with boundaries or other users.
\begin{table}[t!]
\captionsetup{justification=centering}
\caption{MRC model hyperparameters: $\gamma$ and $\lambda$ are discount factor and Temporal-Difference (TD) parameter in reinforcement learning\cite{cohen2021use}.}
\label{tab:hyperparameter}
\scriptsize
\renewcommand{\arraystretch}{1.5}
\centering
\begin{tabular}{|c|c|}
\hline
Hyperparameters                       & Value             \\ \hline
\hline
batch size                           & 2048              \\ \hline
learning rate                        & 0.001             \\ \hline
layers                               & 4                 \\ \hline
hidden units                         & 256               \\ \hline
$\gamma$ (discount factor)           & 0.99              \\ \hline
$\lambda$ (Temporal Difference (TD) 
 learning's parameter) & 0.95              \\ \hline
max steps                            & $1.5 \times 10^7$ \\ \hline
\end{tabular}

\end{table}
\subsection{Reward function}
In a MARL structure, we can set a common goal for multiple users to achieve using the reward function. Therefore, defining a reward function that is appropriate for the goal is important when designing a reinforcement learning model. The reward function R for MRC is as follows:
    \begin{equation}
        R(s_t, a_t) = w_rR_r + w_dR_d + w_aR_a,
    \end{equation}
where $R$ consists of three sub-reward functions, $R_r$, $R_d$, and $R_a$, and the weight $w$ of each reward is determined by several training trials. $R_r$ ($<0$) is the "reset reward" and is computed as a negative value, so that the user receives a penalty each time they resets. The $R_d$ ($>$0) is a positive reward that offsets the penalty and is calculated as the distance traveled from the time $t$ to the next reset. $R_a$ ($>$0) is another positive reward that offsets the penalty and is calculated as the ratio of the area of the visible region \cite{lee1983visibility} in the reset direction of the user's location to the sum of the areas of the visible regions in all resettable directions at time $t$.

$R_r$ is given as a negative value, which is always a constant $-1$, regardless of the current state. If users perform frequent resets during an episode, they will incur a large penalty due to $R_r$. MRC learns the optimal reset direction that results in fewer resets while incurring as little penalty as possible, thus maximizing the cumulative reward. 

Suppose the index of the user who performed a reset at time $t$ is $j$. If $d_j$ is the distance the $j$-th user walked in virtual space from the time of the reset at $t$ to the time of the reset again, then the function for $R_d$ is as follows:
    \begin{equation}
        R_d(s_t, a_t) = d_j,
    \end{equation}
where $R_d$ is computed as a larger reward if the $j$-th user walks further after the reset and $d_j$ becomes larger. MRC learns a reset direction that allows the user to walk further from the location at time $t$ by $R_d$. As users experiencing RDW walk longer distances between resets, the frequency of resets decreases during the experience.

$R_a$ is another positive reward that offsets the penalty. The $f(\theta)$ used to calculate $R_a$ is the distance from the user's position to the nearest wall, obstacle, or other user in the $\theta$ direction. Integrating this over all resettable directions gives the total visible area, which is the sum of the areas of the visible regions in all resettable directions from the user's position. On the other hand, the forward visible area is computed as the sum of the area of the visible region over the direction vectors of the region bounded by $\frac{\pi}{8}$ to the left and right of the reset direction $\theta_a$ determined by MRC, rather than all possible reset directions. The value $\frac{\pi}{8}$ is an appropriate value for training, obtained from several simulations. Let $R_a$ be the fraction of the total visible area that is taken up by the forward visible area, which is calculated as follows:
    \begin{equation}
    \begin{aligned}
        R_a(s_t, a_t) & = \frac {\mbox{forward~visible~area}} {\mbox{total~visible~area}} \\[2pt]
        & = \frac {\displaystyle\int_{\theta_{a}-{\pi/8}}^{\theta_{a}+{\pi/8}} {1 \over 2} (f(\theta))^2 d\theta} {\displaystyle\int_{-\pi}^{\pi} {1 \over 2} (f(\theta))^2 d\theta},
    \end{aligned}
    \end{equation}
where, a larger forward visible area means that the user's reset direction is further away from obstacles and other users, leaving a larger walkable area. With $R_a$, MRC learns to reset the user in a direction with a larger walkable area. This prevents frequent resets due to small walkable areas and helps MRC to learn the reset direction, especially in complex physical spaces with many obstacles.

\subsection{Training Environment}
We trained our model on an Intel i7-12700 CPU, NVIDIA RTX 2080 Ti, and 64GB of RAM. The hyperparameters used for training are listed in \cref{tab:hyperparameter}. The simulation environment was implemented at 30 frames per second. We applied redirection gains to users based on detection thresholds measured in previous studies \cite{steinicke2009estimation, steinicke2009real, cho2021walking, williams2019estimation, li2021detection, langbehn2017bending}. Therefore, we applied thresholds of $7.5$m, $[0.86, 1.26]$, and $[0.67, 1.24]$ for curvature, translation, and rotation gains, respectively, to the users. In the simulation we set the forward and rotational speeds of the user to 1.4 m/s and $90~^{\circ}$/s respectively. 

\section{Evaluation}
\begin{figure}[t!]
    \centering
    \begin{subfigure}[b]{0.3\columnwidth}
  	\centering
  	\includegraphics[width=\textwidth]{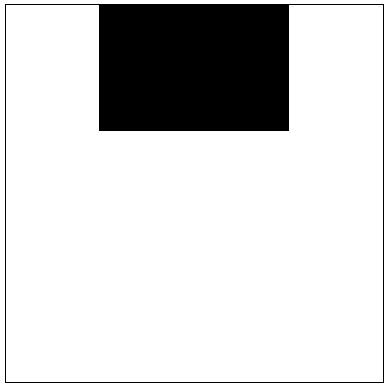}
  	\caption{Simple}
  	\label{4a}
    \end{subfigure}%
    \begin{subfigure}[b]{0.3\columnwidth}
  	\centering
  	\includegraphics[width=\textwidth]{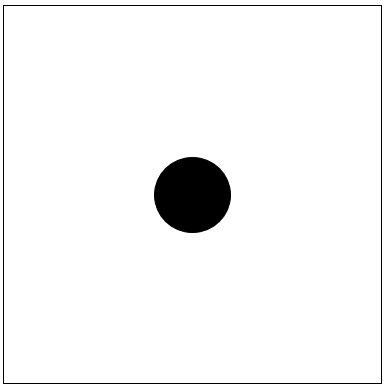}
  	\caption{Circle}
  	\label{4b}
    \end{subfigure}%
    \begin{subfigure}[b]{0.3\columnwidth}
  	\centering
  	\includegraphics[width=\textwidth]{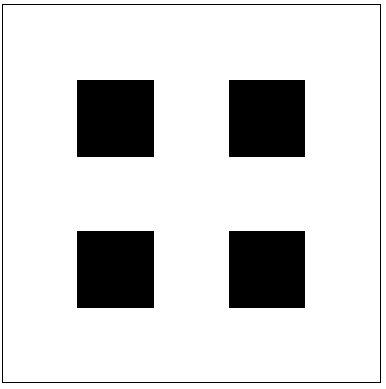}
  	\caption{Four squares}
  	\label{4c}
    \end{subfigure}%
    \\
    \begin{subfigure}[b]{0.3\columnwidth}
  	\centering
  	\includegraphics[width=\textwidth]{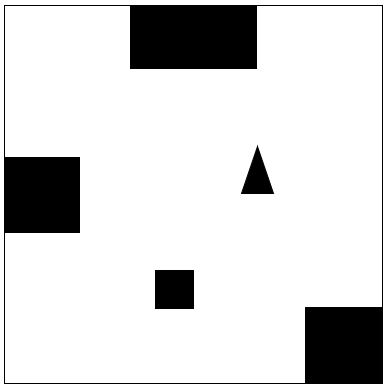}
  	\caption{Complex}
  	\label{4d}
    \end{subfigure}%
    \begin{subfigure}[b]{0.3\columnwidth}
  	\centering
  	\includegraphics[width=\textwidth]{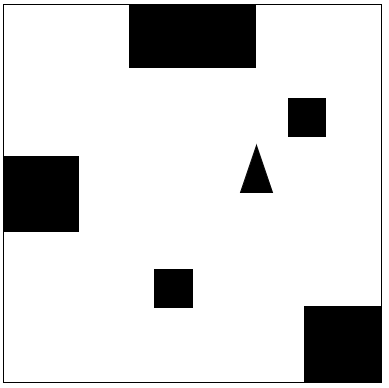}
  	\caption{More}
  	\label{4e}
    \end{subfigure}%
    \begin{subfigure}[b]{0.3\columnwidth}
  	\centering
  	\includegraphics[width=\textwidth]{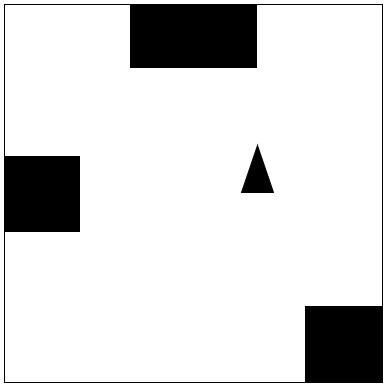}
  	\caption{Less}
  	\label{4f}
    \end{subfigure}%
    \subfigsCaption{
    Types of obstacles in the physical space used for simulation and user study.}
    \label{fig:obs}
\end{figure}

Our experiments compare three redirection algorithms, APF \cite{thomas2019general}, S2C \cite{razzaque2005redirected}, and NS (No-Steering: applying no redirection), with three reset controllers, MRC, R2C, and R2G \cite{thomas2019general}. All controllers except R2C use the Open-RDW \cite{li2021openrdw} and R2C is implemented as described by Thomas et al.~\cite{thomas2019general}. We experiment with six different physical spaces, shown in \cref{fig:obs}, to confirm how MRC performs in physical spaces with different shapes and numbers of obstacles. In \cref{4a}, a simple rectangular obstacle is placed at the top center of the physical space. In \cref{4b}, a circle-shaped obstacle is placed in the center of the physical space. In \cref{4c}, four square obstacles are placed between the center and the corners of the physical space. In \cref{4d} different shaped obstacles are placed in a complex in the physical space. In \cref{4f} and \cref{4e} a new obstacle is added to the complex type or an existing obstacle is removed, respectively. 

\subsection{Simulation Test}

\subsubsection{Simulation Test Design}
We first evaluated the performance of MRC through simulations. The settings for the size of the physical space, the type of obstacle, and the number of simulated users for each experiment are shown below:
\begin{itemize}
    \item E1: $5 \times 5$ m$^2$ / simple, circle, complex / 2 users.
    \item E2: $10 \times 10$ m$^2$ / simple, circle, complex / 2-3 users.
    \item E3: $20 \times 20$ m$^2$ / four squares / 2-8 users.
    \item E4: $10 \times 10$ m$^2$ / less, more / 2 users.
    \item E5: $10 \times 10$ m$^2$ /simple / 2 users.
\end{itemize}
We trained MRC used in our experiments individually based on the size of the physical space, the type of obstacle, and the redirection algorithm. In all simulation experiments, the virtual space in which users walked was a square of $100 \times 100$ m$^2$. We measured the number of resets by walking a random path 100 times for each condition. We set the number of resets as the dependent variable and analyzed the effect of the reset technique, the type of obstacle, and the number of users on the dependent variable. Since the data measured in the experiment did not satisfy the normality and equality of variance through the Shapiro-Wilk and Levene tests, the statistical test was performed by the N-way ANOVA test after aligned rank transform \cite{wobbrock2011aligned} or Kruskal-Wallis test. We also used the Mann-Whitney U test as a post-hoc test. 

\begin{figure}[t!]
    \centering
        \includegraphics[width=\linewidth]{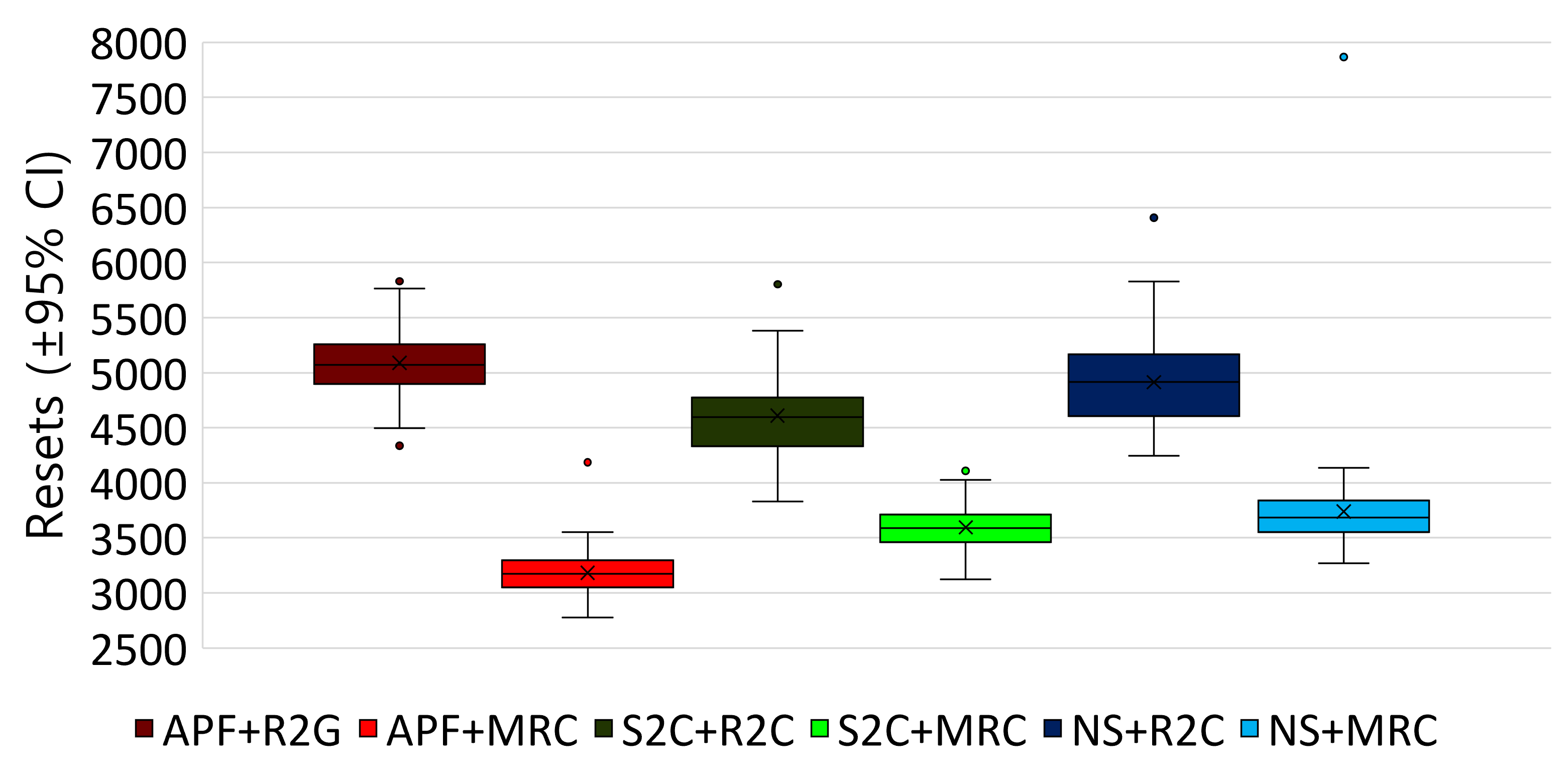} 
        \caption{Visualization of the results from E1 conducted by two users in $5 \times 5$ m$^2$ physical space with complex obstacles}
        \label{fig:5chart}
\end{figure}

\begin{table*}[t!]
\captionsetup{justification=centering}
\caption{The result for each obstacle type and redirection controller of E1, the mean and standard deviation of the number of resets during the simulation with two users, and the results of Mann-Whitney U test with each pair of redirection algorithm and reset technique. $^{***}$ means that the p-value is less than $0.001$.}
\label{tab:5_2_resets}
\scriptsize
\centering
\renewcommand{\arraystretch}{1.5}
\resizebox*{\textwidth}{!}{%
\begin{tabular}{|c|cccccc|ccc|}
\hline
Type of&
  \multicolumn{6}{c|}{Redirection algorithm + Reset technique} &
  \multicolumn{3}{c|}{Statistics result} \\ \cline{2-10} 
obsacles &
  \multicolumn{1}{c|}{APF + R2G} &
  \multicolumn{1}{c|}{S2C + R2C} &
  \multicolumn{1}{c|}{NS + R2C} &
  \multicolumn{1}{c|}{APF + MRC} &
  \multicolumn{1}{c|}{S2C + MRC} &
  NS + MRC &
  \multicolumn{1}{c|}{APF} &
  \multicolumn{1}{c|}{S2C} &
  NS \\ \hline \hline
Simple &
  \multicolumn{1}{c|}{\textit{\begin{tabular}[c]{@{}c@{}}M = 2992.80\\      SD = 167.15\end{tabular}}} &
  \multicolumn{1}{c|}{\textit{\begin{tabular}[c]{@{}c@{}}M = 3208.53\\      SD = 192.04\end{tabular}}} &
  \multicolumn{1}{c|}{\textit{\begin{tabular}[c]{@{}c@{}}M = 3272.53\\      SD = 214.08\end{tabular}}} &
  \multicolumn{1}{c|}{\textit{\begin{tabular}[c]{@{}c@{}}M = 2494.39\\      SD = 127.64\end{tabular}}} &
  \multicolumn{1}{c|}{\textit{\begin{tabular}[c]{@{}c@{}}M = 2482.95\\      SD = 132.88\end{tabular}}} &
  \textit{\begin{tabular}[c]{@{}c@{}}M = 2860.88\\      SD = 189.54\end{tabular}} &
  \multicolumn{1}{c|}{$u(198) = {12.00}^{***}$} &
  \multicolumn{1}{c|}{$u(198) = {12.21}^{***}$} &
  \multicolumn{1}{c|}{$u(198) = {10.95}^{***}$} \\ \hline
Circle &
  \multicolumn{1}{c|}{\textit{\begin{tabular}[c]{@{}c@{}}M = 2419.79\\      SD = 122.76\end{tabular}}} &
  \multicolumn{1}{c|}{\textit{\begin{tabular}[c]{@{}c@{}}M = 5887.89\\      SD = 351.59\end{tabular}}} &
  \multicolumn{1}{c|}{\textit{\begin{tabular}[c]{@{}c@{}}M = 5936.53\\      SD = 541.26\end{tabular}}} &
  \multicolumn{1}{c|}{\textit{\begin{tabular}[c]{@{}c@{}}M = 2329.95\\      SD = 127.83\end{tabular}}} &
  \multicolumn{1}{c|}{\textit{\begin{tabular}[c]{@{}c@{}}M = 2594.55\\      SD = 135.85\end{tabular}}} &
  \textit{\begin{tabular}[c]{@{}c@{}}M = 2583.25\\      SD = 134.10\end{tabular}} &
  \multicolumn{1}{c|}{$u(198) = {4.66}^{***}$} &
  \multicolumn{1}{c|}{$u(198) = {12.22}^{***}$} &
  \multicolumn{1}{c|}{$u(198) = {12.22}^{***}$} \\ \hline
Complex &
  \multicolumn{1}{c|}{\textit{\begin{tabular}[c]{@{}c@{}}M = 5090.14\\      SD = 293.33\end{tabular}}} &
  \multicolumn{1}{c|}{\textit{\begin{tabular}[c]{@{}c@{}}M = 4909.68\\      SD = 335.35\end{tabular}}} &
  \multicolumn{1}{c|}{\textit{\begin{tabular}[c]{@{}c@{}}M = 4913.57\\      SD = 377.83\end{tabular}}} &
  \multicolumn{1}{c|}{\textit{\begin{tabular}[c]{@{}c@{}}M = 3182.18\\      SD = 196.37\end{tabular}}} &
  \multicolumn{1}{c|}{\textit{\begin{tabular}[c]{@{}c@{}}M = 3596.05\\      SD = 185.36\end{tabular}}} &
  \textit{\begin{tabular}[c]{@{}c@{}}M = 3739.23\\      SD = 459.30\end{tabular}} &
  \multicolumn{1}{c|}{$u(198) = {12.22}^{***}$} &
  \multicolumn{1}{c|}{$u(198) = {12.16}^{***}$} &
  \multicolumn{1}{c|}{$u(198) = {12.22}^{***}$} \\ \hline
\end{tabular}%
}
\end{table*}
\begin{table*}[t!]
\captionsetup{justification=centering}
\caption{The result for each obstacle type and redirection controller of E2, the mean and standard deviation of the number of resets during the simulation with two users, and the results of Mann-Whitney U test with each pair of redirection algorithm and reset technique. $^{***}$ means that the p-value is less than $0.001$.}
\label{tab:10_2_resets}
\renewcommand{\arraystretch}{1.5}
\centering
\resizebox{\textwidth}{!}{%
\begin{tabular}{|c|cccccc|ccc|}
\hline
Type of &
  \multicolumn{6}{c|}{Redirection algorithm + Reset technique} &
  \multicolumn{3}{c|}{Statistics result} \\ \cline{2-10} 
obstacles &
  \multicolumn{1}{c|}{APF + R2G} &
  \multicolumn{1}{c|}{S2C + R2C} &
  \multicolumn{1}{c|}{NS + R2C} &
  \multicolumn{1}{c|}{APF + MRC} &
  \multicolumn{1}{c|}{S2C + MRC} &
  NS + MRC &
  \multicolumn{1}{c|}{APF} &
  \multicolumn{1}{c|}{S2C} &
  NS \\ \hline \hline
Simple &
  \multicolumn{1}{c|}{\textit{\begin{tabular}[c]{@{}c@{}}M = 1647.49\\      SD = 92.38\end{tabular}}} &
  \multicolumn{1}{c|}{\textit{\begin{tabular}[c]{@{}c@{}}M = 1828.65\\      SD = 121.06\end{tabular}}} &
  \multicolumn{1}{c|}{\textit{\begin{tabular}[c]{@{}c@{}}M = 1824.16\\      SD = 110.70\end{tabular}}} &
  \multicolumn{1}{c|}{\textit{\begin{tabular}[c]{@{}c@{}}M = 1269.21\\      SD = 68.61\end{tabular}}} &
  \multicolumn{1}{c|}{\textit{\begin{tabular}[c]{@{}c@{}}M = 1395.43\\      SD = 77.48\end{tabular}}} &
  \textit{\begin{tabular}[c]{@{}c@{}}M = 1486.43\\      SD = 79.62\end{tabular}} &
  \multicolumn{1}{c|}{$u(198) = {12.21}^{***}$} &
  \multicolumn{1}{c|}{$u(198) = {12.21}^{***}$} &
  \multicolumn{1}{c|}{$u(198) = {12.02}^{***}$} \\ \hline
Circle &
  \multicolumn{1}{c|}{\textit{\begin{tabular}[c]{@{}c@{}}M = 1516.47\\      SD = 84.88\end{tabular}}} &
  \multicolumn{1}{c|}{\textit{\begin{tabular}[c]{@{}c@{}}M = 2920.99\\      SD = 165.42\end{tabular}}} &
  \multicolumn{1}{c|}{\textit{\begin{tabular}[c]{@{}c@{}}M = 3054.09\\      SD = 196.65\end{tabular}}} &
  \multicolumn{1}{c|}{\textit{\begin{tabular}[c]{@{}c@{}}M = 1273.91\\      SD = 71.09\end{tabular}}} &
  \multicolumn{1}{c|}{\textit{\begin{tabular}[c]{@{}c@{}}M = 2077.53\\      SD = 119.46\end{tabular}}} &
  \textit{\begin{tabular}[c]{@{}c@{}}M = 1467.71\\      SD = 76.92\end{tabular}} &
  \multicolumn{1}{c|}{$u(198) = {11.89}^{***}$} &
  \multicolumn{1}{c|}{$u(198) = {12.22}^{***}$} &
  \multicolumn{1}{c|}{$u(198) = {12.22}^{***}$} \\ \hline
Complex &
  \multicolumn{1}{c|}{\textit{\begin{tabular}[c]{@{}c@{}}M = 2736.20\\      SD = 156.64\end{tabular}}} &
  \multicolumn{1}{c|}{\textit{\begin{tabular}[c]{@{}c@{}}M = 2900.36\\      SD = 223.52\end{tabular}}} &
  \multicolumn{1}{c|}{\textit{\begin{tabular}[c]{@{}c@{}}M = 2956.97\\      SD = 264.93\end{tabular}}} &
  \multicolumn{1}{c|}{\textit{\begin{tabular}[c]{@{}c@{}}M = 1955.79\\      SD = 111.82\end{tabular}}} &
  \multicolumn{1}{c|}{\textit{\begin{tabular}[c]{@{}c@{}}M = 2101.49\\      SD = 130.32\end{tabular}}} &
  \textit{\begin{tabular}[c]{@{}c@{}}M = 2089.88\\      SD = 105.28\end{tabular}} &
  \multicolumn{1}{c|}{$u(198) = {12.22}^{***}$} &
  \multicolumn{1}{c|}{$u(198) = {12.18}^{***}$} &
  \multicolumn{1}{c|}{$u(198) = {12.22}^{***}$} \\ \hline
\end{tabular} }
\end{table*}
\begin{table*}[t!]
\captionsetup{justification=centering}
\caption{The result for each obstacle type and redirection controller of E2, the mean and standard deviation of the number of resets during the simulation with three users, and the results of Mann-Whitney U test with each pair of redirection algorithm and reset technique. $^{***}$ means that the p-value is less than $0.001$.}
\label{tab:10_3_resets}
\renewcommand{\arraystretch}{1.5}
\centering
\resizebox{\textwidth}{!}{%
\begin{tabular}{|c|cccccc|ccc|}
\hline
Type of &
  \multicolumn{6}{c|}{Redirection algorithm + Reset technique} &
  \multicolumn{3}{c|}{Statistics result} \\ \cline{2-10} 
obstacles &
  \multicolumn{1}{c|}{APF + R2G} &
  \multicolumn{1}{c|}{S2C + R2C} &
  \multicolumn{1}{c|}{NS + R2C} &
  \multicolumn{1}{c|}{APF + MRC} &
  \multicolumn{1}{c|}{S2C + MRC} &
  NS + MRC &
  \multicolumn{1}{c|}{APF} &
  \multicolumn{1}{c|}{S2C} &
  NS \\ \hline \hline
Simple &
  \multicolumn{1}{c|}{\textit{\begin{tabular}[c]{@{}c@{}}M = 2236.09\\      SD = 135.21\end{tabular}}} &
  \multicolumn{1}{c|}{\textit{\begin{tabular}[c]{@{}c@{}}M = 3390.90\\      SD = 229.90\end{tabular}}} &
  \multicolumn{1}{c|}{\textit{\begin{tabular}[c]{@{}c@{}}M = 3369.95\\      SD = 204.63\end{tabular}}} &
  \multicolumn{1}{c|}{\textit{\begin{tabular}[c]{@{}c@{}}M = 2236.09\\      SD = 135.21\end{tabular}}} &
  \multicolumn{1}{c|}{\textit{\begin{tabular}[c]{@{}c@{}}M = 2497.50\\      SD = 138.94\end{tabular}}} &
  \textit{\begin{tabular}[c]{@{}c@{}}M = 2610.24\\      SD = 139.49\end{tabular}} &
  \multicolumn{1}{c|}{$u(198) = {12.13}^{***}$} &
  \multicolumn{1}{c|}{$u(198) = {12.21}^{***}$} &
  \multicolumn{1}{c|}{$u(198) = {12.21}^{***}$} \\ \hline
Circle &
  \multicolumn{1}{c|}{\textit{\begin{tabular}[c]{@{}c@{}}M = 2461.18\\      SD = 126.17\end{tabular}}} &
  \multicolumn{1}{c|}{\textit{\begin{tabular}[c]{@{}c@{}}M = 4503.56\\      SD = 252.54\end{tabular}}} &
  \multicolumn{1}{c|}{\textit{\begin{tabular}[c]{@{}c@{}}M = 5412.22\\      SD = 469.77\end{tabular}}} &
  \multicolumn{1}{c|}{\textit{\begin{tabular}[c]{@{}c@{}}M = 2202.37\\      SD = 118.35\end{tabular}}} &
  \multicolumn{1}{c|}{\textit{\begin{tabular}[c]{@{}c@{}}M = 3389.61\\      SD = 195.95\end{tabular}}} &
  \textit{\begin{tabular}[c]{@{}c@{}}M = 2501.38\\      SD = 130.14\end{tabular}} &
  \multicolumn{1}{c|}{$u(198) = {10.56}^{***}$} &
  \multicolumn{1}{c|}{$u(198) = {12.22}^{***}$} &
  \multicolumn{1}{c|}{$u(198) = {12.22}^{***}$} \\ \hline
Complex &
  \multicolumn{1}{c|}{\textit{\begin{tabular}[c]{@{}c@{}}M = 4291.60\\      SD = 241.13\end{tabular}}} &
  \multicolumn{1}{c|}{\textit{\begin{tabular}[c]{@{}c@{}}M = 5061.79\\      SD = 407.18\end{tabular}}} &
  \multicolumn{1}{c|}{\textit{\begin{tabular}[c]{@{}c@{}}M = 5412.22\\      SD = 469.77\end{tabular}}} &
  \multicolumn{1}{c|}{\textit{\begin{tabular}[c]{@{}c@{}}M = 3398.90\\      SD = 216.61\end{tabular}}} &
  \multicolumn{1}{c|}{\textit{\begin{tabular}[c]{@{}c@{}}M = 3729.05\\      SD = 229.53\end{tabular}}} &
  \textit{\begin{tabular}[c]{@{}c@{}}M = 3572.22\\      SD = 191.61\end{tabular}} &
  \multicolumn{1}{c|}{$u(198) = {11.96}^{***}$} &
  \multicolumn{1}{c|}{$u(198) = {12.21}^{***}$} &
  \multicolumn{1}{c|}{$u(198) = {12.22}^{***}$} \\ \hline
\end{tabular} }
\end{table*}

\subsubsection{Result of E1: $5 \times 5$ m$^2$ Physical Space}
\cref{tab:5_2_resets} shows the results of E1. In a physical space with a circle type obstacle, MRC applied to the NS algorithm for two users reduced the mean number of resets by 56.49\% compared to R2C. We performed a 3 (type of obstacles) $\times$ 2 (reset technique) two-way ANOVA test to determine the effect of the type of obstacles and reset technique on the number of resets for each redirection algorithm. The type of obstacles and the reset technique each had a large effect on the number of resets (detailed statistical results in supplementary document's Table 1). The interaction of the two independent variables showed a medium effect when using the APF algorithm and a large effect for S2C and NS. We then performed a post-hoc Mann-Whitney U test to confirm that MRC significantly reduced the mean number of resets compared to the other reset technique in all conditions ($<.001$, detailed statistical result in \cref{tab:5_2_resets}). \cref{fig:5chart} shows the performance of each redirection controller in physical space with complex type obstacles. Through E1, we confirmed that MRC performs well compared to existing reset techniques in a small-sized physical space, even in the presence of different types of obstacles.

\begin{figure}[t!]
    \centering
        \includegraphics[width=\linewidth]{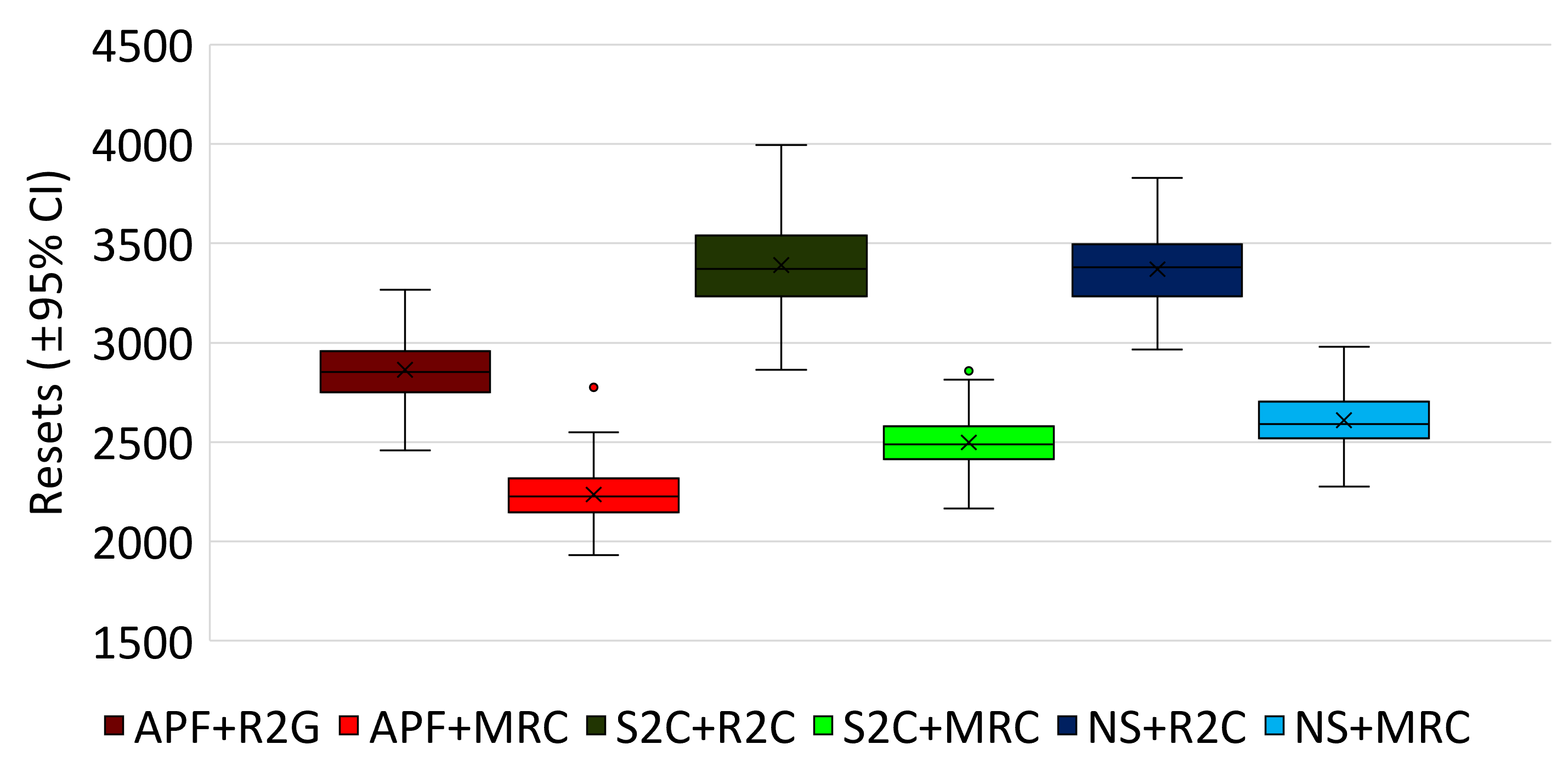} 
        \caption{Visualization of the results from E2 conducted by three users in $10 \times 10$ m$^2$ physical space with simple obstacles}
        \label{fig:10chart}
\end{figure}

\begin{figure}[t!]
    \centering
        \includegraphics[width=\linewidth,height=0.5\linewidth]{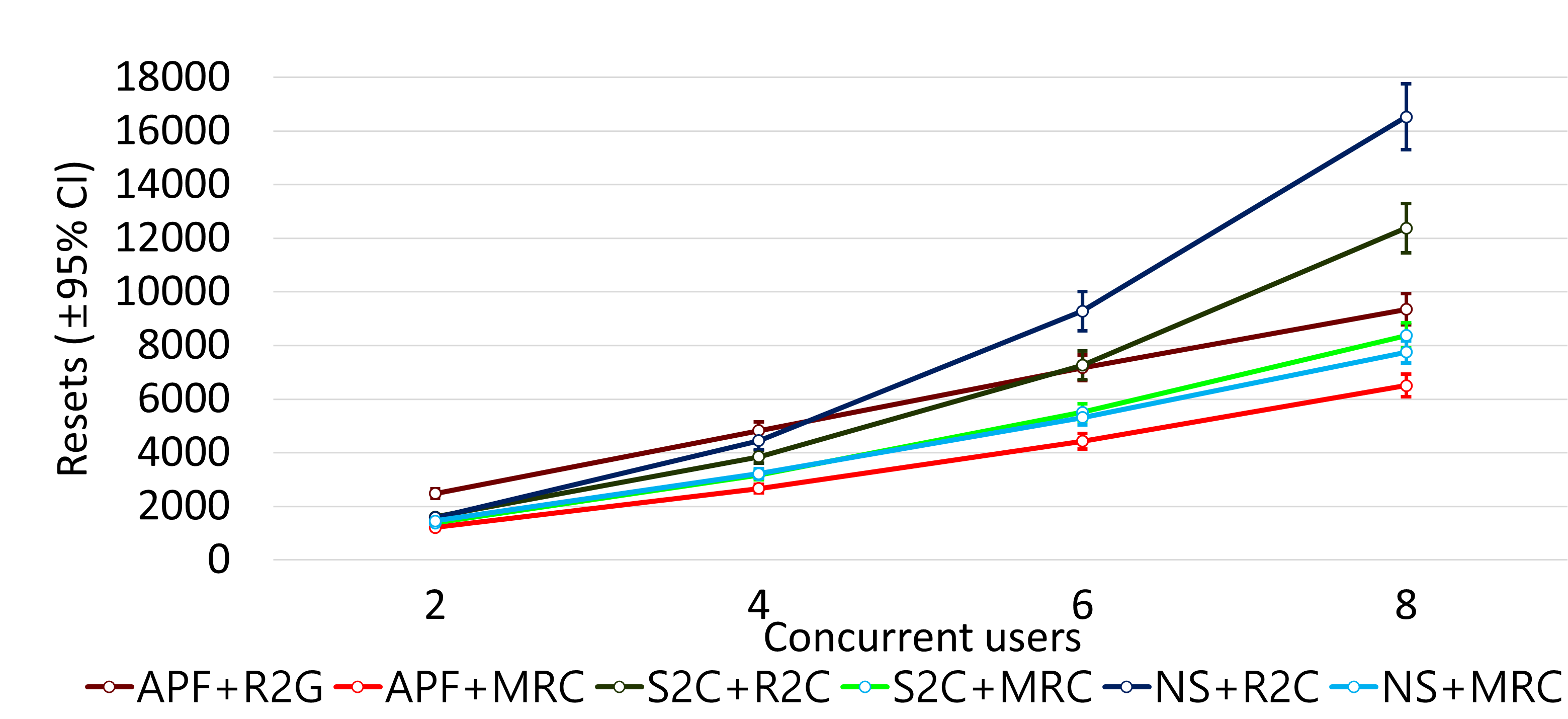} 
        \caption{Visualization of the results from E3 conducted by 2-8 users in $20 \times 20$ m$^2$ physical space with four square obstacles}
        \label{fig:20chart}
\end{figure}

\subsubsection{Result of E2: $10 \times 10$ m$^2$ Physical Space}
The results of the simulations with two and three users in E2 are shown in \cref{tab:10_2_resets} and \cref{tab:10_3_resets}, respectively. As shown in E1, applying MRC to the NS algorithm for three users in a physical space with a circle type obstacle reduced the mean number of resets by 53.78\% compared to R2C. We performed a 3 (type of obstacles) $\times$ 2 (reset technique) two-way ANOVA test to determine the effect of the type of obstacles and the reset technique on the number of resets when using each redirection algorithm for two and three users, respectively. The type of obstacles and the reset technique had large effects on the number of resets (detailed statistical results in supplementary document's Table 2 and Table 3). The interaction of the two independent variables had a medium effect on the number of resets when using the S2C algorithm, and a large effect when using APF and NS. We then performed a post-hoc Mann-Whitney U test to confirm that MRC significantly reduced the mean number of resets compared to the other reset technique in all conditions ($<.001$, detailed statistical result in \cref{tab:10_3_resets} and \cref{tab:10_3_resets}). \cref{fig:10chart} shows the performance of each redirection controller for three users in the physical space with a simple type obstacle. Through E2, we confirmed that MRC performs well compared to existing reset techniques in a medium-sized physical space, even in the presence of different types of obstacles.
\begin{figure}[t!]
    \centering
        
        \includegraphics[width=\linewidth]{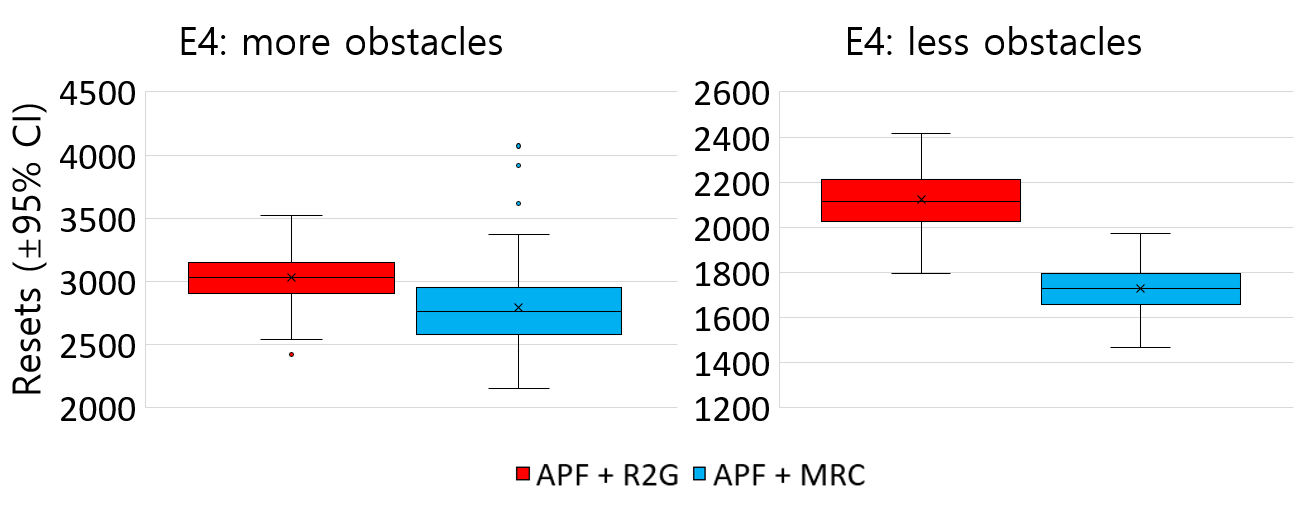} 
        \caption{Visualization of the results of E4 conducted by two users applying the APF algorithm in $10 \times 10$ m$^2$ physical space with less and more obstacle types.}
        \label{fig:unseen_chart}
\end{figure}
\subsubsection{Result of E3: $20 \times 20$ m$^2$ Physical Space}
\cref{fig:20chart} shows the performance of each redirection controller based on simulations with two to eight users in E3. We performed a 4 (the number of users) $\times$ 2 (reset technique) two-way ANOVA test to determine the effect of the number of users and the reset technique on the number of resets for each redirection algorithm. The number of users and the reset techniques had a large effect on the number of resets (detailed statistical results in supplementary document's Table 4). The interaction of the two independent variables had a large effect when using APF and NS algorithms. However, when the S2C algorithm was used, the interaction had a small effect. S2C redirects all users to the center, and as the number of users increases, the number of resets increases due to frequent resets in the center of the physical space. Although, we change the reset technique, frequent resets will still occur in the center, so the interaction had small effect on the number of resets. We then performed a post-hoc Mann-Whitney U test to confirm that MRC significantly reduced the mean number of resets compared to the other reset technique in all conditions ($<.001$, detailed statistical results in supplementary document's Table 5). Through E3, we confirmed that MRC performs well compared to existing reset techniques in the large-sized physical space, even in situations where varying numbers of users experience RDW.

\subsubsection{Result of E4: Unseen Type of Obstacles}
In E4, we confirmed how MRC performed in the physical space of the same size as that experienced during training, but with a similarly different type of obstacle. We trained MRC, which would be used in E4, in $10 \times 10$ m$^2$ physical space with a complex type of obstacle. We set up the physical spaces in E4 by adding or removing an obstacle to the physical space where MRC was trained. We performed the simulations with two users by applying R2G and MRC to the APF algorithm. We performed a 2 (type of obstacles) $\times$ 2 (reset technique) two-way ANOVA test to determine the effect of the type of obstacles and the reset technique on the number of resets. The type of obstacle and the reset technique had a large effect on the number of resets, and their interaction had a medium effect (detailed statistical results in supplementary document's Table 6 and Table 7). We then performed a post-hoc Mann-Whitney U test to confirm that MRC significantly reduced the mean number of resets compared to R2G in both environments (less: $u(198) = 11.99^{***}$\footnote{\label{p3} *** means that p-value is less than 0.001.}, more: $u(198) = 6.22^{***}$\footref{p3}). \cref{fig:unseen_chart} shows the performance of applying R2G and MRC to the APF algorithm in E4. Through E4, we have confirmed that MRC performs better than existing reset techniques in the physical spaces with similar types of obstacles to the trained environment.

\subsubsection{Result of E5: Unseen Size of Physical Space}
In contrast to E4, in E5 we confirmed how MRC performed in the physical space of the same type of obstacle, but with a different size of physical space. We trained MRC\_5 and MRC\_10 for use in E5 on a 5$x 5$ m$^2$ and a 10$x 10$ m$^2$ physical space, respectively, in which a simple type obstacle was placed. The simulations with two users were performed in the physical space where MRC\_10 was trained by applying R2G, MRC\_5, and MRC\_10 to the APF algorithm. We performed a Kruskal-Wallis test to determine the effect of reset technique on the number of resets. The reset technique had a large effect on the number of resets ($\chi^2 = 215.38^{***}\footref{p3}, \eta^2 = 0.72$). A post-hoc Mann-Whitney U test confirmed that MRC\_5 had a significantly lower mean number of resets than R2G ($u(297) = 10.02^{***}$\footref{p3}, detailed results of the mean and standard deviation of the number of resets in supplementary document's Table 8). \cref{fig:diff_chart} shows the performance of each reset technique of the APF algorithm in E5. Through E5, we have confirmed that MRC outperforms the existing reset technique, even when applied to an environment of a different size than a trained MRC. The fact that MRC was trained using normalized information from physical spaces enabled it to perform well in the environments of different sizes.
\begin{figure}[t!]
    \centering
        \includegraphics[width=\linewidth]{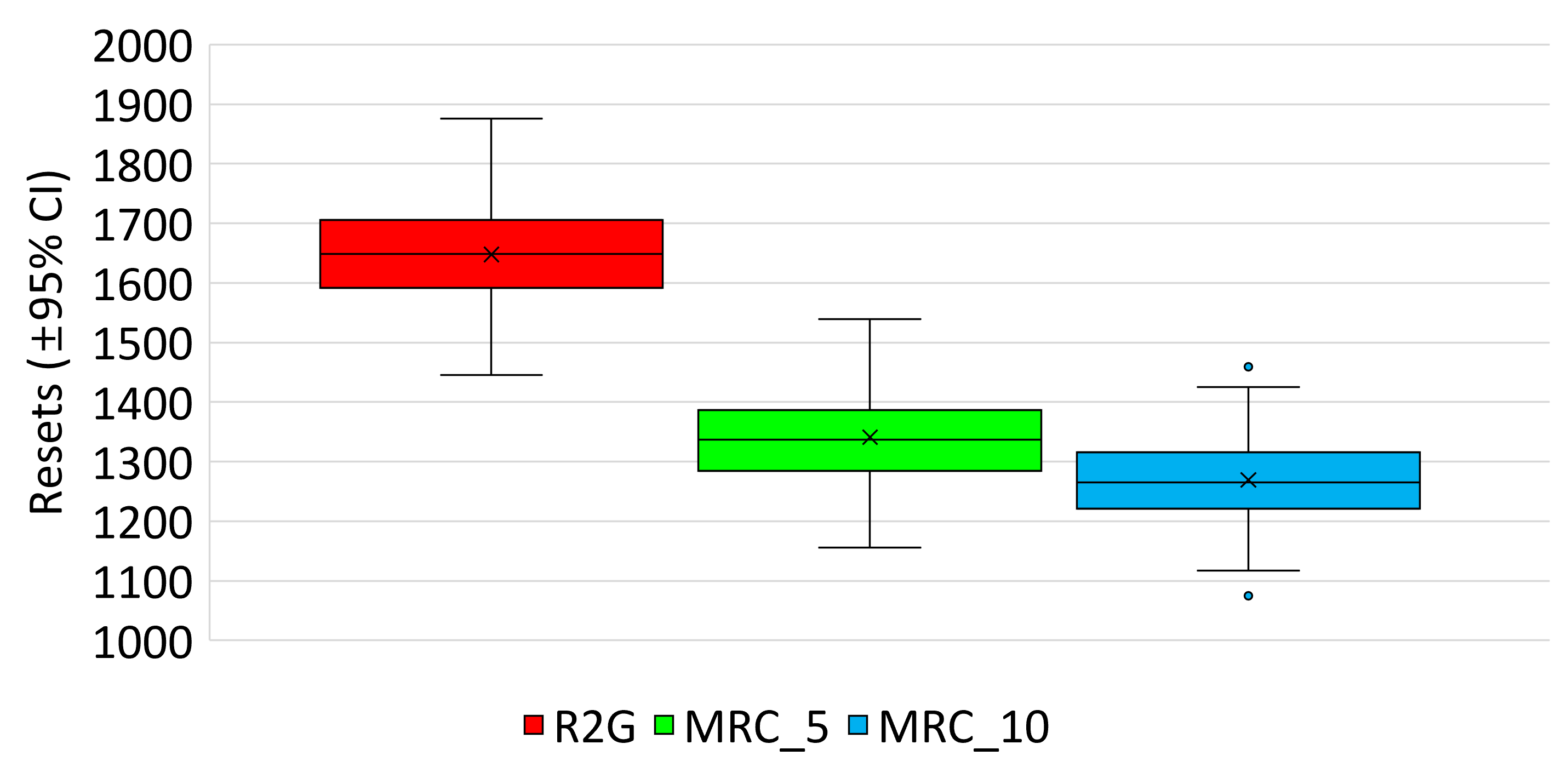} 
        \caption{Visualization of the results from E5 conducted by two users in $10 \times 10$ m$^2$ physical space with simple obstacles}
        \label{fig:diff_chart}
\end{figure}
\subsection{User Study}
\subsubsection{User Study Design}
Following the simulation test, we conducted a user study to verify how MRC performed in a real multi-user environment. We used HTC VIVE Cosmos Elite HMDs to render and display each virtual scene to the user. In order for the HMD to remotely communicate with our desktops and measure each user's location and orientation, we placed four base stations in the corners of the physical space. We conducted a user study in a physical space of $6 \times 6$ m$^2$ with obstacles of complex type (\cref{4d}). We did not install actual obstacles, but instead performed a reset when a user reached the boundary of an obstacle, to prevent the user from being injured by collisions with obstacles.

\begin{figure}[t!]
    \centering
        \begin{subfigure}[b]{0.8\columnwidth}
      	\centering
      	\includegraphics[width=\textwidth,height=0.2\textheight]{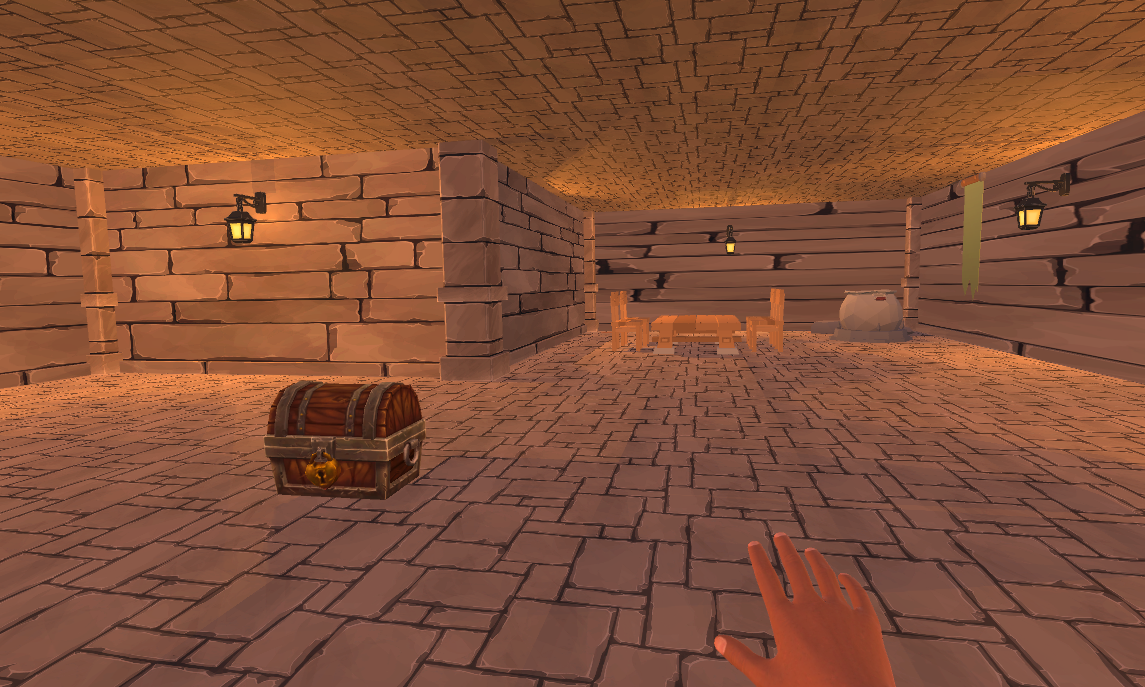}
            \caption{user's view while walking in virtual environment.}
      	\label{10a}
        \end{subfigure}%
        \hspace{2mm}
        \begin{subfigure}[b]{0.8\columnwidth}
      	\centering
      	\includegraphics[width=\textwidth,height=0.2\textheight]{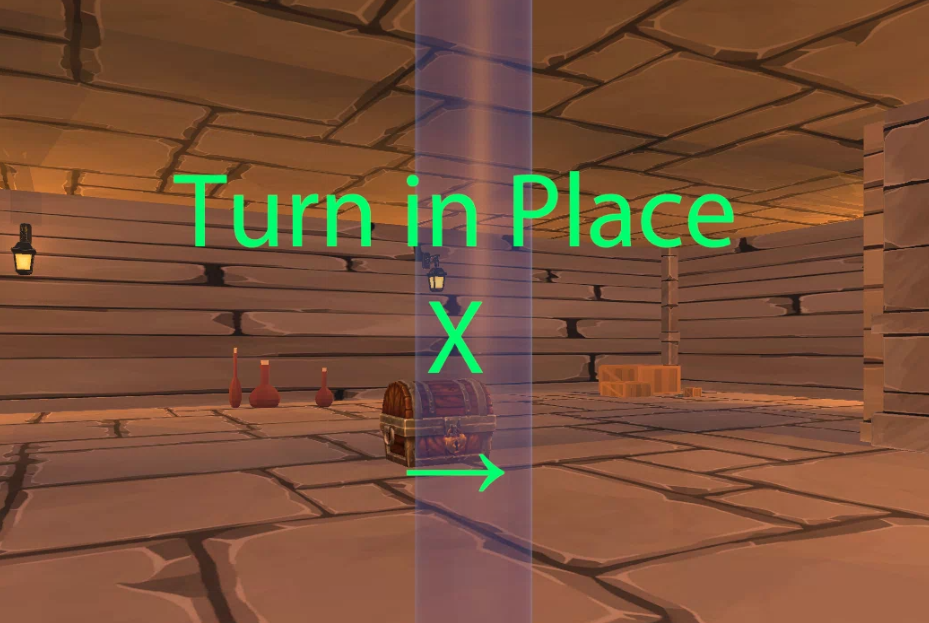}
            \caption{user's view while resetting.}
      	\label{10b}
        \end{subfigure}%
    \subfigsCaption{The virtual environment's scenes of our user study.}
    \label{fig:user test}
\end{figure}

As with the simulation experiment, we implemented our user study experiment based on the Open-RDW \cite{li2021openrdw}. In the experiment, users explored a virtual dungeon-like environment of $40 \times 40$ m$^2$ in order to find the treasure chests. The controller held by the user was rendered as a hand in the virtual environment. \cref{10a} shows the view of a user exploring a virtual space. If the user pressed the trigger of the controller near the treasure chest, the hand would be rendered as a grabbing motion and the treasure chest would disappear. A new treasure chest would then be randomly generated 4-8 meters away from the user's current location. Users followed a randomized path through the dungeon to find randomly generated treasures. To enhance the user experience, we played sound effects such as wind noise on the HMD device during the experiment, and played an alarm sound when a treasure chest was disappeared. Some users unfamiliar with RDW rotated more than the direction indicated by the reset controller. In our experiments, we used a novel UI display to help redirect the user in the right direction. When the user was in a reset situation, the word "Turn in Place", an arrow indicating the direction to turn, and an 'X' in the center of the screen were generated on the user's HMD screen. In addition, a transparent cylinder was generated in front of the user inside the dungeon. If the user rotated in place until the transparent cylinder and the 'X' icon were aligned, the UI text would disappear. After the text disappeared, the cylinder would disappear 0.5 seconds later. Using this interface, the user could perform a reset exactly as the reset controller indicated, and easily find the direction they were walking in virtual space before the reset. \cref{10b} shows the UI display as seen on the HMD screen of the user performing the reset.

We applied MRC and the existing reset techniques (R2C, R2G) \cite{thomas2019general} to the APF algorithm to compare how they performed in our user study. We counterbalanced the order of the reset techniques applied to each participant by dividing them into six groups to eliminate carryover effects. Each participant experienced RDW for 90 seconds with each reset technique. In our experiment, we measured the number of resets and the mean distance between resets in the virtual environment (MDbR) to prepare our dataset. The G$^*$power $3.1$ program\cite{faul2007g} was used to determine the minimum size for an experiment with three reset techniques as independent variables to be statistically significant. We needed a sample size of 28 participants to satisfy the significance level ($= 0.05$), medium effect size ($= 0.25$), and statistical power ($= 0.8$). We recruited participants by advertising that we would pay them 8 US dollars (\$) to participate in an RDW experiment. The recruited participants were organized into teams of two and experienced RDW sharing a physical space.

We also used four questionnaires to measure the participants' experience of motion sickness, immersion, and presence during the user study. The four questionnaires are SSQ \cite{kennedy1993simulator} (for motion sickness), E2I \cite{lin2002effects} (for immersion), SUSPQ \cite{usoh2000using}, and IPQ \cite{schubert2001experience} (for presence). Before starting the VR experience, participants were asked to complete a pre-SSQ questionnaire. The experiment is repeated three times, with participants experiencing 90 seconds of VR, followed by 10 minutes of questionnaire completion and a break.
\subsubsection{Result of User Study}
A total of 28 people (male:female $= 19:9$) participated in the experiment. The age of the participants was between $20-36$ ($M = 23.79, SD = 8.03$). $14$ participants had tried one to two VR experiences, seven had tried three or more, and the rest had never tried VR. The mean of the Pre-SSQ scored $6.68~(SD = 17.37)$.

\begin{figure}[t!]
    \centering
        
        \includegraphics[width=\linewidth]{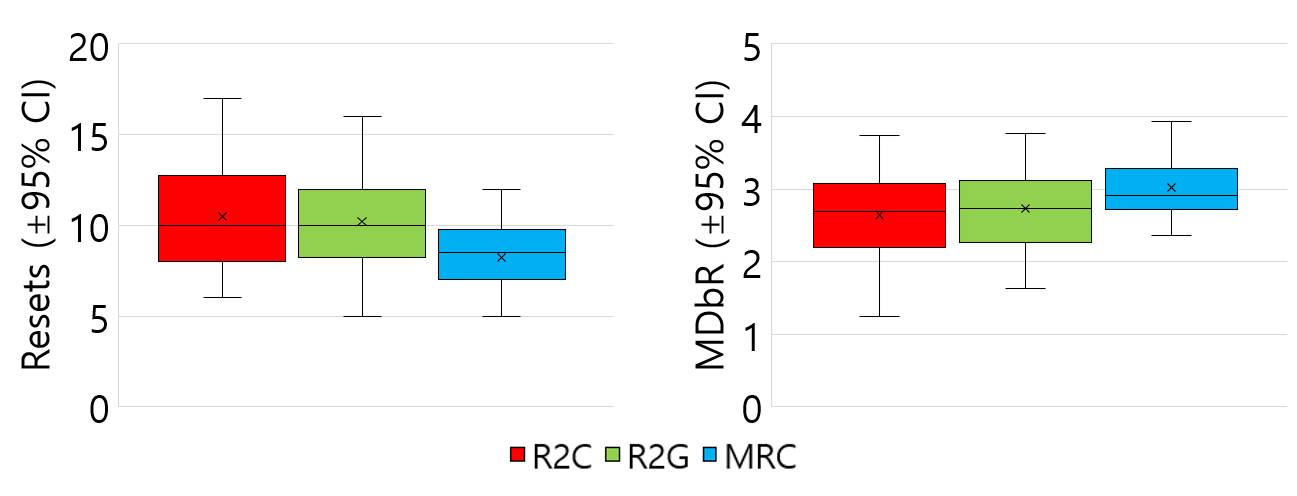} 
        \caption{Visualization of the number of resets and the MDBR for each reset technique in a user study.}
        \label{fig:user_chart}
\end{figure}

\cref{fig:user_chart} shows the performance by the number of resets and by the MDbR of applying each reset technique in the user study. We performed a one-way repeated measures ANOVA (RM-ANOVA) because the number of resets data passed the pre-test. The reset technique had a large effect on the number of resets ($F(2,54) = 10.98^{***}\footref{p3}, \eta^2 = 0.14$). We then performed a post-hoc t-test with Bonferroni correction to confirm that MRC had a significantly lower mean number of resets than R2C and R2G ($p < p_c = .017$). The MDbR data did not pass the pre-test, so a Friedman test was performed. We found significant differences in the MDbR results for each reset technique ($\chi^2 = 9.21, p < .01$). We then performed a post-hoc Wilcoxon signed-rank test with Bonferroni correction and confirmed that MRC had a significantly greater MDbR than R2C and R2G ($p < p_c = .017$). All questionnaire results showed no significant differences between the reset techniques. We expect to confirm significant differences in the questionnaire results as we have more experience with RDW over a longer period of time and see greater differences in the number of resets for each reset technique. Detailed results of the mean and standard deviation, and visualization of each condition are presented in Table 9 and Figure 1 of the supplementary document. Through the user study, we confirmed that MRC outperforms the existing reset techniques by resetting the user in the optimal direction, even when applied to a real multi-user environment.
\section{MARL Evaluation}
We conducted an experiment to confirm that learning MRC using the MARL method performs better than using Single Agent Reinforcement Learning (SARL). We use one of the SARL methods, the SAC \cite{haarnoja2018soft}, which was used to train OSP \cite{jeon2022dynamic}. Assuming that MRC\_S is a multi-user reset controller trained with SAC, the MDP for training it is almost identical to the original MRC. However, MRC\_S allows a single actor to observe the physical space information of all users and determine the reset direction for each user individually. Assuming that the number of users experiencing RDW is $n$, the dimension of the state that an actor in MRC\_S must observe and the dimension of the action space that an actor in MRC\_S must determine are $n$ times larger than those of the original MRC.

MRC\_S was trained as an RDW experience episode with two users applying the APF algorithm in a $10 \times 10$ m$^2$ physical space with a complex type obstacle (\cref{4d}). Then, we applied MRC\_S to the APF algorithm for two users and measured the number of resets in 100 experiments in the same virtual environment and path as the simulation test. With MRC\_S, the mean number of resets is 33\% higher than with original MRC. We statistically analyzed the effect of R2G, MRC, and MRC\_S on the number of resets. A one-way ANOVA test was performed as the data passed the pre-test. The reset technique had a large effect on the number of resets ($F(2,297) = 955.20^{***}\footref{p3}, \eta^2 = 0.87$). We then performed a post hoc Tukey HSD test and confirmed that MRC\_S significantly reduced the mean number of rests compared to R2G. However, we also confirmed that MRC\_S significantly increased the mean number of resets compared to original MRC ($p < .001$). \cref{tab:ablstudy} shows the mean and standard deviation of the number of resets for each reset technique. Through a MARL evaluation, we confirmed that the SARL-trained MRC did not learn the globally optimized reset direction.

\section{Discussion}
Compared to the existing reset techniques, we confirmed that MRC reduces the number of resets regardless of the size of the physical space, the type of obstacles, and the number of users. In addition, MRC compensates for the weaknesses of each of the redirection algorithms. For example, the APF algorithm is generally superior to other methods in most environments because it redirects users in the direction of a sum of forces calculated in real time from the user's location. However, when there are many obstacles of complex type in a small physical space such as E1, we can observe that APF actually performs worse than simple S2C. This is because various forces from nearby obstacles constantly disrupt the redirection of the APF with a limited range of redirection gains. However, when MRC was applied to the APF, it performed better than when it was applied to the S2C, which can be explained as MRC helping APF redirect users. 
\begin{table}[t!]
\captionsetup{justification=centering}
\caption{The result for the MARL evaluation, mean and standard deviation of the number of resets for each RDW controller: MRC\_S is a model trained with the SARL method instead of MARL.}
\label{tab:ablstudy}
\renewcommand{\arraystretch}{1.5}
\centering
\begin{tabular}{|c|ccc|}
\hline
Result & \multicolumn{3}{c|}{Redirection algorithm + Reset technique}                  \\ \cline{2-4} 
                        & \multicolumn{1}{c|}{APF + R2G} & \multicolumn{1}{c|}{APF + MRC\_S} & APF + MRC \\ \hline \hline
{$M$}              & \multicolumn{1}{c|}{2736.2}    & \multicolumn{1}{c|}{2612.86}     & 1955.79   \\ \hline
\textit{SD}             & \multicolumn{1}{c|}{156.6401}  & \multicolumn{1}{c|}{135.0389}    & 111.8229  \\ \hline
\end{tabular}
\end{table}
MRC models trained by reinforcement learning generally perform well in other physical spaces with similar characteristics to the space used for training because they use normalized spatial information as data. This allows a trained MRC to be used in a physical space similar to the trained space without additional training.

Despite its many advantages, MRC has some limitations. First of all, in order to make the most effective use of MRC, it is necessary to train the MRC model separately for each particular environment. Otherwise, MRC may not perform well in environments with obstacles that are differ from the MRC training environment. We also tried to reset users in the optimal direction with MRC, but confirmed that users unfamiliar with RDW had difficulty rotating correctly in the reset direction. Therefore, research is needed on how to help users perform resets in the correct direction.

\section{Conclusions}
We propose MRC, a reset controller that can be effectively used when dynamically moving multi-users share a physical space with various obstacles to experience RDW. MRC significantly reduced the number of resets during experiences compared to existing reset techniques in a multi-user environment under various conditions (number of users, types of obstacles, redirection algorithms). In addition, we found that using MARL to train MRC outperformed than using SARL in the MARL evaluation. We expect to be able to solve problems in various multi-user RDW environments using the MARL algorithm in the future. 

We propose as future work to study reset controllers that use different types of MARL methods to reduce learning time and achieve better performance. There are many variations of MARL methods, including the MA-POCA used in MRC. We expect to be able to compare different methods to find the best MARL methods for multi-user RDW environments. We also propose research into different ways to address the problem that users unfamiliar with the RDW may not perform resets exactly in the direction determined by the MRC. It would be very interesting to design different ways to naturally help users perform a reset in the correct direction and see the effect on reset performance.

\acknowledgments{
	This research was supported by the National Research Foundation of Korea (No. NRF-2020R1A2C2014622, No. NRF-2022R1I1A1A01072228) and Korea Radio Promotion Association (No. RNIX20230200) grant funded by the Korea government (MSIT, MSE).
}

\bibliography{Paper.bbl}

\begin{thebibliography}{10}

\bibitem{azmandian2015physical}
M.~Azmandian, T.~Grechkin, M.~T. Bolas, and E.~A. Suma.
\newblock Physical space requirements for redirected walking: How size and
  shape affect performance.
\newblock In {\em ICAT-EGVE}, pp. 93--100, 2015.

\bibitem{azmandian2017evaluation}
M.~Azmandian, T.~Grechkin, and E.~S. Rosenberg.
\newblock An evaluation of strategies for two-user redirected walking in shared
  physical spaces.
\newblock In {\em 2017 IEEE Virtual Reality (VR)}, pp. 91--98. IEEE, 2017.

\bibitem{bachmann2019multi}
E.~R. Bachmann, E.~Hodgson, C.~Hoffbauer, and J.~Messinger.
\newblock Multi-user redirected walking and resetting using artificial
  potential fields.
\newblock {\em IEEE transactions on visualization and computer graphics},
  25(5):2022--2031, 2019.

\bibitem{bucsoniu2010multi}
L.~Bu{\c{s}}oniu, R.~Babu{\v{s}}ka, and B.~D. Schutter.
\newblock Multi-agent reinforcement learning: An overview.
\newblock {\em Innovations in multi-agent systems and applications-1}, pp.
  183--221, 2010.

\bibitem{chang2021redirection}
Y.~Chang, K.~Matsumoto, T.~Narumi, T.~Tanikawa, and M.~Hirose.
\newblock Redirection controller using reinforcement learning.
\newblock {\em IEEE Access}, 9:145083--145097, 2021.

\bibitem{cho2021walking}
Y.-H. Cho, D.-H. Min, J.-S. Huh, S.-H. Lee, J.-S. Yoon, and I.-K. Lee.
\newblock Walking outside the box: Estimation of detection thresholds for
  non-forward steps.
\newblock In {\em 2021 IEEE Virtual Reality and 3D User Interfaces (VR)}, pp.
  448--454. IEEE, 2021.

\bibitem{cohen2021use}
A.~Cohen, E.~Teng, V.-P. Berges, R.-P. Dong, H.~Henry, M.~Mattar, A.~Zook, and
  S.~Ganguly.
\newblock On the use and misuse of absorbing states in multi-agent
  reinforcement learning.
\newblock {\em arXiv preprint arXiv:2111.05992}, 2021.

\bibitem{faul2007g}
F.~Faul, E.~Erdfelder, A.-G. Lang, and A.~Buchner.
\newblock G* power 3: A flexible statistical power analysis program for the
  social, behavioral, and biomedical sciences.
\newblock {\em Behavior research methods}, 39(2):175--191, 2007.

\bibitem{haarnoja2018soft}
T.~Haarnoja, A.~Zhou, P.~Abbeel, and S.~Levine.
\newblock Soft actor-critic: Off-policy maximum entropy deep reinforcement
  learning with a stochastic actor.
\newblock In {\em International conference on machine learning}, pp.
  1861--1870. PMLR, 2018.

\bibitem{jeon2022dynamic}
S.-B. Jeon, S.-U. Kwon, J.-Y. Hwang, Y.-H. Cho, H.~Kim, J.~Park, and I.-K. Lee.
\newblock Dynamic optimal space partitioning for redirected walking in
  multi-user environment.
\newblock {\em ACM Transactions on Graphics (TOG)}, 41(4):1--14, 2022.

\bibitem{juliani2018unity}
A.~Juliani, V.-P. Berges, E.~Teng, A.~Cohen, J.~Harper, C.~Elion, C.~Goy,
  Y.~Gao, H.~Henry, M.~Mattar, et~al.
\newblock Unity: A general platform for intelligent agents.
\newblock {\em arXiv preprint arXiv:1809.02627}, 2018.

\bibitem{kennedy1993simulator}
R.~S. Kennedy, N.~E. Lane, K.~S. Berbaum, and M.~G. Lilienthal.
\newblock Simulator sickness questionnaire: An enhanced method for quantifying
  simulator sickness.
\newblock {\em The international journal of aviation psychology},
  3(3):203--220, 1993.

\bibitem{kim2021adjusting}
D.~Kim, J.-e. Shin, J.~Lee, and W.~Woo.
\newblock Adjusting relative translation gains according to space size in
  redirected walking for mixed reality mutual space generation.
\newblock In {\em 2021 IEEE Virtual Reality and 3D User Interfaces (VR)}, pp.
  653--660. IEEE, 2021.

\bibitem{langbehn2017bending}
E.~Langbehn, P.~Lubos, G.~Bruder, and F.~Steinicke.
\newblock Bending the curve: Sensitivity to bending of curved paths and
  application in room-scale vr.
\newblock {\em IEEE transactions on visualization and computer graphics},
  23(4):1389--1398, 2017.

\bibitem{langbehn2018evaluation}
E.~Langbehn, P.~Lubos, and F.~Steinicke.
\newblock Evaluation of locomotion techniques for room-scale vr: Joystick,
  teleportation, and redirected walking.
\newblock In {\em Proceedings of the Virtual Reality International
  Conference-Laval Virtual}, pp. 1--9, 2018.

\bibitem{lee1983visibility}
D.-T. Lee.
\newblock Visibility of a simple polygon.
\newblock {\em Computer Vision, Graphics, and Image Processing},
  22(2):207--221, 1983.

\bibitem{lee2019real}
D.-Y. Lee, Y.-H. Cho, and I.-K. Lee.
\newblock Real-time optimal planning for redirected walking using deep
  q-learning.
\newblock In {\em 2019 IEEE Conference on Virtual Reality and 3D User
  Interfaces (VR)}, pp. 63--71. IEEE, 2019.

\bibitem{lee2020optimal}
D.-Y. Lee, Y.-H. Cho, D.-H. Min, and I.-K. Lee.
\newblock Optimal planning for redirected walking based on reinforcement
  learning in multi-user environment with irregularly shaped physical space.
\newblock In {\em 2020 IEEE Conference on Virtual Reality and 3D User
  Interfaces (VR)}, pp. 155--163. IEEE, 2020.

\bibitem{li2021detection}
Y.-J. Li, D.-R. Jin, M.~Wang, J.-L. Chen, F.~Steinicke, S.-M. Hu, and Q.~Zhao.
\newblock Detection thresholds with joint horizontal and vertical gains in
  redirected jumping.
\newblock In {\em 2021 IEEE Virtual Reality and 3D User Interfaces (VR)}, pp.
  95--102. IEEE, 2021.

\bibitem{li2021openrdw}
Y.-J. Li, M.~Wang, F.~Steinicke, and Q.~Zhao.
\newblock Openrdw: A redirected walking library and benchmark with multi-user,
  learning-based functionalities and state-of-the-art algorithms.
\newblock In {\em 2021 IEEE International Symposium on Mixed and Augmented
  Reality (ISMAR)}, pp. 21--30. IEEE, 2021.

\bibitem{lin2002effects}
J.-W. Lin, H.~B.-L. Duh, D.~E. Parker, H.~Abi-Rached, and T.~A. Furness.
\newblock Effects of field of view on presence, enjoyment, memory, and
  simulator sickness in a virtual environment.
\newblock In {\em Proceedings ieee virtual reality 2002}, pp. 164--171. IEEE,
  2002.

\bibitem{lowe2017multi}
R.~Lowe, Y.~I. Wu, A.~Tamar, J.~Harb, O.~Pieter~Abbeel, and I.~Mordatch.
\newblock Multi-agent actor-critic for mixed cooperative-competitive
  environments.
\newblock {\em Advances in neural information processing systems}, 30, 2017.

\bibitem{messinger2019effects}
J.~Messinger, E.~Hodgson, and E.~R. Bachmann.
\newblock Effects of tracking area shape and size on artificial potential field
  redirected walking.
\newblock In {\em 2019 IEEE Conference on Virtual Reality and 3D User
  Interfaces (VR)}, pp. 72--80. IEEE, 2019.

\bibitem{nescher2014planning}
T.~Nescher, Y.-Y. Huang, and A.~Kunz.
\newblock Planning redirection techniques for optimal free walking experience
  using model predictive control.
\newblock In {\em 2014 IEEE Symposium on 3D User Interfaces (3DUI)}, pp.
  111--118. IEEE, 2014.

\bibitem{nilsson201815}
N.~C. Nilsson, T.~Peck, G.~Bruder, E.~Hodgson, S.~Serafin, M.~Whitton,
  F.~Steinicke, and E.~S. Rosenberg.
\newblock 15 years of research on redirected walking in immersive virtual
  environments.
\newblock {\em IEEE computer graphics and applications}, 38(2):44--56, 2018.

\bibitem{5759437}
T.~C. Peck, H.~Fuchs, and M.~C. Whitton.
\newblock An evaluation of navigational ability comparing redirected free
  exploration with distractors to walking-in-place and joystick locomotio
  interfaces.
\newblock In {\em 2011 IEEE Virtual Reality Conference}, pp. 55--62, 2011.
  \href{https://doi.org/10.1109/VR.2011.5759437}
{doi: {{%
10\hspace{.1pt}\discretionary{.}{%
}{.}\hspace{.4pt}1109\discretionary{/}{%
}{/}VR\hspace{.1pt}\discretionary{.}{%
}{.}\hspace{.4pt}2011\hspace{.1pt}\discretionary{.}{%
}{.}\hspace{.4pt}5759437}}}


\bibitem{puterman2014markov}
M.~L. Puterman.
\newblock {\em Markov decision processes: discrete stochastic dynamic
  programming}.
\newblock John Wiley \& Sons, 2014.

\bibitem{razzaque2005redirected}
S.~Razzaque.
\newblock {\em Redirected walking}.
\newblock The University of North Carolina at Chapel Hill, 2005.

\bibitem{schubert2001experience}
T.~Schubert, F.~Friedmann, and H.~Regenbrecht.
\newblock The experience of presence: Factor analytic insights.
\newblock {\em Presence: Teleoperators \& Virtual Environments},
  10(3):266--281, 2001.

\bibitem{schulman2017proximal}
J.~Schulman, F.~Wolski, P.~Dhariwal, A.~Radford, and O.~Klimov.
\newblock Proximal policy optimization algorithms.
\newblock {\em arXiv preprint arXiv:1707.06347}, 2017.

\bibitem{steinicke2009real}
F.~Steinicke, G.~Bruder, K.~Hinrichs, J.~Jerald, H.~Frenz, and M.~Lappe.
\newblock Real walking through virtual environments by redirection techniques.
\newblock {\em JVRB-Journal of Virtual Reality and Broadcasting}, 6(2), 2009.

\bibitem{steinicke2009estimation}
F.~Steinicke, G.~Bruder, J.~Jerald, H.~Frenz, and M.~Lappe.
\newblock Estimation of detection thresholds for redirected walking techniques.
\newblock {\em IEEE transactions on visualization and computer graphics},
  16(1):17--27, 2009.

\bibitem{strauss2020steering}
R.~R. Strauss, R.~Ramanujan, A.~Becker, and T.~C. Peck.
\newblock A steering algorithm for redirected walking using reinforcement
  learning.
\newblock {\em IEEE transactions on visualization and computer graphics},
  26(5):1955--1963, 2020.

\bibitem{suma2012taxonomy}
E.~A. Suma, G.~Bruder, F.~Steinicke, D.~M. Krum, and M.~Bolas.
\newblock A taxonomy for deploying redirection techniques in immersive virtual
  environments.
\newblock In {\em 2012 IEEE Virtual Reality Workshops (VRW)}, pp. 43--46. IEEE,
  2012.

\bibitem{sutton1998introduction}
R.~S. Sutton, A.~G. Barto, et~al.
\newblock {\em Introduction to reinforcement learning}, vol. 135.
\newblock MIT press Cambridge, 1998.

\bibitem{thomas2019general}
J.~Thomas and E.~S. Rosenberg.
\newblock A general reactive algorithm for redirected walking using artificial
  potential functions.
\newblock In {\em 2019 IEEE Conference on Virtual Reality and 3D User
  Interfaces (VR)}, pp. 56--62. IEEE, 2019.

\bibitem{usoh1999walking}
M.~Usoh, K.~Arthur, M.~C. Whitton, R.~Bastos, A.~Steed, M.~Slater, and F.~P.
  Brooks~Jr.
\newblock Walking> walking-in-place> flying, in virtual environments.
\newblock In {\em Proceedings of the 26th annual conference on Computer
  graphics and interactive techniques}, pp. 359--364, 1999.

\bibitem{usoh2000using}
M.~Usoh, E.~Catena, S.~Arman, and M.~Slater.
\newblock Using presence questionnaires in reality.
\newblock {\em Presence}, 9(5):497--503, 2000.

\bibitem{vaswani2017attention}
A.~Vaswani, N.~Shazeer, N.~Parmar, J.~Uszkoreit, L.~Jones, A.~N. Gomez,
  {\L}.~Kaiser, and I.~Polosukhin.
\newblock Attention is all you need.
\newblock {\em Advances in neural information processing systems}, 30, 2017.

\bibitem{williams2007exploring}
B.~Williams, G.~Narasimham, B.~Rump, T.~P. McNamara, T.~H. Carr, J.~Rieser, and
  B.~Bodenheimer.
\newblock Exploring large virtual environments with an hmd when physical space
  is limited.
\newblock In {\em Proceedings of the 4th symposium on Applied perception in
  graphics and visualization}, pp. 41--48, 2007.

\bibitem{williams2021arc}
N.~L. Williams, A.~Bera, and D.~Manocha.
\newblock Arc: Alignment-based redirection controller for redirected walking in
  complex environments.
\newblock {\em IEEE Transactions on Visualization and Computer Graphics},
  27(5):2535--2544, 2021.

\bibitem{williams2019estimation}
N.~L. Williams and T.~C. Peck.
\newblock Estimation of rotation gain thresholds considering fov, gender, and
  distractors.
\newblock {\em IEEE transactions on visualization and computer graphics},
  25(11):3158--3168, 2019.

\bibitem{wobbrock2011aligned}
J.~O. Wobbrock, L.~Findlater, D.~Gergle, and J.~J. Higgins.
\newblock The aligned rank transform for nonparametric factorial analyses using
  only anova procedures.
\newblock In {\em Proceedings of the SIGCHI conference on human factors in
  computing systems}, pp. 143--146, 2011.

\bibitem{9881577}
S.-Z. Xu, T.-Q. Liu, J.-H. Liu, S.~Zollmann, and S.-H. Zhang.
\newblock Making resets away from targets: Poi aware redirected walking.
\newblock {\em IEEE Transactions on Visualization and Computer Graphics},
  28(11):3778--3787, 2022. \href{https://doi.org/10.1109/TVCG.2022.3203095}
{doi: {{%
10\hspace{.1pt}\discretionary{.}{%
}{.}\hspace{.4pt}1109\discretionary{/}{%
}{/}TVCG\hspace{.1pt}\discretionary{.}{%
}{.}\hspace{.4pt}2022\hspace{.1pt}\discretionary{.}{%
}{.}\hspace{.4pt}3203095}}}


\bibitem{zhang2022adaptive}
S.-H. Zhang, C.-H. Chen, Z.~Fu, Y.~Yang, and S.-M. Hu.
\newblock Adaptive optimization algorithm for resetting techniques in
  obstacle-ridden environments.
\newblock {\em IEEE Transactions on Visualization and Computer Graphics}, 2022.

\bibitem{9733261}
S.-H. Zhang, C.-H. Chen, and S.~Zollmann.
\newblock One-step out-of-place resetting for redirected walking in vr.
\newblock {\em IEEE Transactions on Visualization and Computer Graphics}, pp.
  1--1, 2022. \href{https://doi.org/10.1109/TVCG.2022.3158609}
{doi: {{%
10\hspace{.1pt}\discretionary{.}{%
}{.}\hspace{.4pt}1109\discretionary{/}{%
}{/}TVCG\hspace{.1pt}\discretionary{.}{%
}{.}\hspace{.4pt}2022\hspace{.1pt}\discretionary{.}{%
}{.}\hspace{.4pt}3158609}}}


\bibitem{zmuda2013optimizing}
M.~A. Zmuda, J.~L. Wonser, E.~R. Bachmann, and E.~Hodgson.
\newblock Optimizing constrained-environment redirected walking instructions
  using search techniques.
\newblock {\em IEEE transactions on visualization and computer graphics},
  19(11):1872--1884, 2013.

\end{thebibliography}

\end{document}